\documentclass[a4paper,fleqn,usenatbib]{mnras}

\usepackage{newtxtext,newtxmath}
\usepackage[T1]{fontenc}
\usepackage{ae,aecompl}

\usepackage{graphicx}	% Including figure files
\usepackage{amsmath}	% Advanced maths commands
\usepackage{amssymb}	% Extra maths symbols

\newcommand{\revone}{}
\newcommand{\revtwo}{}
\newcommand{\revthree}{}

\title[Observational Evidence For Constant Gas Accretion Rate Since z = 5]{Observational Evidence For Constant Gas Accretion Rate Since z = 5}

\author[E. F. Spring]{
Eleanor F. Spring$^{1}$\thanks{E-mail: eleanorspring@hotmail.co.uk}
and Micha{\l} J.~Micha{\l}owski$^{1,2}$
\\
$^{1}$Institute for Astronomy, University of Edinburgh, Royal Observatory, Blackford Hill, Edinburgh EH9 3HJ, UK\\
$^{2}$Astronomical Observatory Institute, Faculty of Physics, Adam Mickiewicz University, ul.~S{\l}oneczna 36, 60-286 Pozna{\'n}, Poland
}

\date{Accepted XXX. Received YYY; in original form ZZZ}

\pubyear{2016}

\begin{document}
\label{firstpage}
\pagerange{\pageref{firstpage}--\pageref{lastpage}}
\maketitle

\begin{abstract}
Star formation rate density (SFRD) has not been constant throughout the history of the Universe. The rate at which stars form greatly affects the evolution of the Universe, but the factors which drive SFRD evolution remain  uncertain. There must be sufficient amount of gas to fuel the star formation, either as a reservoir within a galaxy, or as inflow from the intergalactic medium (IGM). This work explores how the gas accretion rate onto galaxies over time has affected star formation rate. We propose a novel method of measuring cosmic gas accretion rate. This involves comparing the {\revone comoving} densities of available {\sc Hi} and $\mbox{H}_2$ gas and the densities of existing stars at different redshifts. We constrained gas accretion until z = 5, and we found that the gas accretion rate density (GARD) is relatively constant in the range from z = 5 to z = 0. This constancy in the GARD is not reflected by the SFRD, which declines significantly between z = 1.0 and z = 0. This work suggests that the decline is not due to a reduction in GARD. 
\end{abstract}


%%%%%%%%%%%%%%%%%%%%%%%%%%%%%%%%%%%%%%%%%%%%%%%%%%

%%%%%%%%%%%%%%%%% BODY OF PAPER %%%%%%%%%%%%%%%%%%

\section{Introduction}
\label{sec:intro}

One of the most important issues of the evolution of the Universe is how galaxies acquire gas which fuels star formation. Numerical galaxy formation models require significant gas inflows from the intergalactic medium (IGM) to fuel star formation~{\revone \citep{Dekel06,Dekel09,Schaye10}}, and indeed the current gas reservoirs in many galaxies are too low to sustain the current level of star formation, even for normal galaxies like the Milky Way~\citep{Draine09}. This is reinforced by simulations of~\citet{LHuillier12}, who found that most galaxies assemble their mass through steady gas accretion, and only most massive galaxies grow predominately due to dramatic merger events. However, such inflow process cannot be studied in details because it is very difficult to directly detect it. Only for a few galaxies was the observational indication of gas inflow obtained~\citep{Martin14,Turner15,Rauch16}, and indirect evidence was reviewed by~\citet{Sancisi08}.

~\citet{Michalowski15, Michalowski16} provide evidence for gas inflow directly fueling star formation. This was achieved by reporting the first 21\,cm line observations for long gamma-ray burst galaxies. These observations implied high levels of {\sc Hi} gas in areas of recent star formation, whereas low molecular gas content was reported for GRB hosts~(\citealt{Hatsukade14, Stanway15, Michalowski16}, but see \citealt{Perley17}). This in turn suggests that star formation is either directly fueled by {\sc Hi} gas, or that there is a very efficient conversion process between {\sc Hi} and $\mbox{H}_2$ underway~\citep{Michalowski15}. 

As reported by~\citet{Sancisi08} within the Milky Way itself, rate of star formation has been very constant over the course of its life. This clearly demonstrates that the gas being used up in star formation is somehow being replaced. Whilst it is obvious that gas accretion must be happening for almost all galaxies, it is very unclear how much is taking place. 

Whilst there is a wide range of theories regarding the evolution of star formation rate, there is little observation based evidence available to study the details of this process. Observations suggest that it is the accretion of metal-poor gas that drives the formation of disc galaxies~\citep{Sanchez14, Sanchez13, Cresci10}. There has been a lot of work done to measure the densities of {\sc Hi} and $\mbox{H}_2$ mass in galaxies at varying redshifts. In order to detect $\mbox{H}_2$ gas, the CO molecule is generally used as a tracer~\citep{Carilli13, Bolatto13}.

The aim of this paper is to provide the first observational measurement of the cosmic gas accretion rate density (GARD) by applying a novel method, and comparing it to the measurement of the star formation rate density (SFRD). A cosmological model with $h = 0.7$, $\Omega_{m}= 0.3$, $\Omega_{\Lambda} = 0.7$ is assumed.

\section{Data}
\label{sec:data}

The compilations of $\rho_{HI}$ and $\rho_{H_{2}}$ from~\citet{Lagos14} and~\citet{Hoppmann15} were used. Significantly more $\rho_{HI}$ data was available than $\rho_{H_{2}}$, over a wider redshift range. We used the compilation of  $\rho_{stellar}$ {\revone (including stellar remnants)} from \citet{Michalowski10}. 

The {\sc Hi} densities were obtained by a variety of methods over the redshift range. These included spectral stacking of the 21-cm Hydrogen emission line and damped Lyman-$\alpha$ (DLA) absorbers. The complete compilation of $\rho_{HI}$ measurements is presented in table~\ref{tab:HIdata}. {\revone All density values are densities per unit of comoving volume.}

The values from~\citet{Zwaan05} and~\citet{Martin10} were for the local universe, at z = 0. The~\citet{Martin10} value was based on the Arecibo Legacy Fast ALFA (ALFALFA) survey, which at that time had completed source extraction for 40\% of the total sky area, allowing the value to be calculated from a sample of 10\,119 galaxies. In comparison, the~\citet{Zwaan05} value was based on the  {\sc Hi} Parkes All Sky Survey (HIPASS) which was based on 4315 extra galactic emission-line detections. As such, it was reasonable to assume that the~\citet{Martin10} value is superior to the~\citet{Zwaan05} value. The~\citet{Freudling11} value was given relative to z = 0, so required a conversion using an established value of $\rho_{HI}$ at z = 0. The~\citet{Lah07} was based on observations from the Giant Metrewave Radio Telescope (GMRT) in order to measure $\rho_{HI}$ at z = 0.24. The~\citet{Delhaize13} values were based on a combination of detected sources and the spectral stacking technique. Contrary to previous estimates, they also suggested that $\rho_{HI}$ evolution over the last 1 Gyr was minimal. The~\citet{Rhee13} values were obtained using {\sc Hi} signal stacking technique. 
The~\citet{Hoppmann15} values also relied upon the detection of hydrogen 21-cm emission with Arecibo Ultra Deep Survey. 

\citet{Chang10} used the {\sc Hi} intensity mapping based on the DEEP2 optical galaxy redshift survey. The~\citet{Masui13} value was obtained by cross-correlating hydrogen 21-cm information with data from the WiggleZ Dark Energy Survey.

The~\citet{Peroux03},~\citet{Rao06},~\citet{Noterdaeme12},~\citet{Zafar13}, and~\citet{Prochaska09} values were all obtained using data from surveys of damped Lyman-$\alpha$ (DLA) absorbers. 

The $\rho_{H_{2}}$ values were sourced from~\citet{Keres03} for z = 0, and~\citet{Decarli16} for other redshifts.

\section{Method: Gas Accretion Rate Density (GARD) Estimate}
\label{sec:method}

{\revtwo In the calculations presented below care must be taken in consistently treating the extent of the gas and stellar components which are considered to lie inside or outside a galaxy. Here we broadly define a galaxy extent as the size of the atomic gas disk, which is usually a few times larger than the stellar disk, i.e.~extends for a few tens of kpc. In this definition the hot gas in the dark matter halo is not considered.}

{\revone Absorption measurements of $\rho_{HI}$ in principle could trace the gas outside galaxies, but the DLA systems used for these studies are dense enough to safely assume that they are associated with galaxies. Indeed, DLAs have SFRs and velocity dispersions characteristic for galaxies, and metallicities higher than that of the IGM \citep[e.g.][]{Moller04,Wolfe05,Ledoux06, Fynbo10,Fynbo11,Fynbo13,Krogager12,Krogager13}. Moreover within the error bars there is no step change from {\sc Hi}-line- to DLA-derived $\rho_{HI}$ values (Fig.~\ref{fig:HI_z} and Table~\ref{tab:HIdata}), which would be expected if DLAs probed additional neutral gas component in the IGM.}
{\revtwo Finally, the sizes of the absorbing gas of DLAs are $\sim10$--$20$\,kpc \citep{Moller04,Peroux05,Wolfe05,Monier09,Fynbo10,Fynbo11,Fynbo13,Krogager12,Krogager13}, safely within our limit of a few tens of kpc. There are some known DLAs with impact parameters from the background quasar of 50--100\,kpc \citep[right part of Fig.~18 in][]{Rao11}, but these DLAs are rare \citep[three out of 27 DLAs in the sample of ][]{Rao11}, so their impact on the measured $\rho_{HI}$ is minor. Moreover, they all have low identification confidence. Similarly, simulations show that the hydrogen density required to classify as a DLA ($2\times10^{20}\,\mbox{cm}^{-2}$; \citealt{Wolfe05}) is only present within $\sim20$\,kpc from the galaxy centre (\citealt{Liang16} fig.~7 and 18; Shen et al., in prep.).}

Calculating the GARD involves the following steps. The total matter density inside galaxies for each redshift bin was calculated as:

\begin{equation}
    \rho_{total} =  \rho_{HI} + \rho_{H_{2}} + \rho_{stel}. 
	\label{eq:rho_total}
\end{equation}

\noindent Over time this density changes only by inflows and outflows:

\begin{equation}
    \rho_{a} = \rho_{b} + \rho_{inflow} - \rho_{outflow}
    \label{eq:rho_a}
\end{equation}

\noindent Where $a$ and $b$ represent total $\rho$ at different epochs. {\revone We do not distinguish different mechanisms of inflow and outflow, so these terms include all processes that add to and remove gas from galaxies, respectively.} {\revtwo The inflow processes include cold and hot mode accretion, as long as the gas ends up inside galaxies (as defined above) during the relevant time span.} {\revthree It also includes gas expelled to the galactic halo during star-formation episodes which is subsequently re-accreted. Galactic fountains eject cold high-metallicity gas into the corona, and by mixing, cool the low-metallicity gas of the corona sufficiently that fountain clouds form and inflow back into the galactic disc \citep{Fraternali14}. Gas accreted from fountain clouds is considered inflow.} Accretion of gas into intra-cluster medium is not included, as such gas does not end up inside galaxies, and is not included in the {\sc Hi} and $H_2$ measurements. {\revone The outflows include supernova and AGN feedback, gas stripping, etc.}

The average density of gas accreted by galaxies between two epochs (i.e. the difference between the gas flowing in and out) can be calculated as:

\begin{equation}
    \mbox{GAD} \equiv \rho_{inflow} - \rho_{outflow} = \rho_{a} - \rho_{b}
    \label{eq:rho_acc}
\end{equation}

\noindent By calculating the \emph{difference} in $\rho_{total}$ between consecutive epochs, {\revone represented in equation~\ref{eq:rho_acc} by $\rho_a$ and $\rho_b$}, only the gas remaining within the galaxy --- either free or as stars --- is accounted for. This quantity was labelled the gas accretion density (GAD). From there, the GARD could then be calculated using:

\begin{equation}
    \mbox{GARD} =\frac{\mbox{GAD}}{\Delta \mbox{time}}.
	\label{eq:GARD}
\end{equation}

\noindent Where $\Delta$time was the time elapsed between neighboring redshift bins.

\section{Results}

\begin{figure}
	\includegraphics[width=\columnwidth]{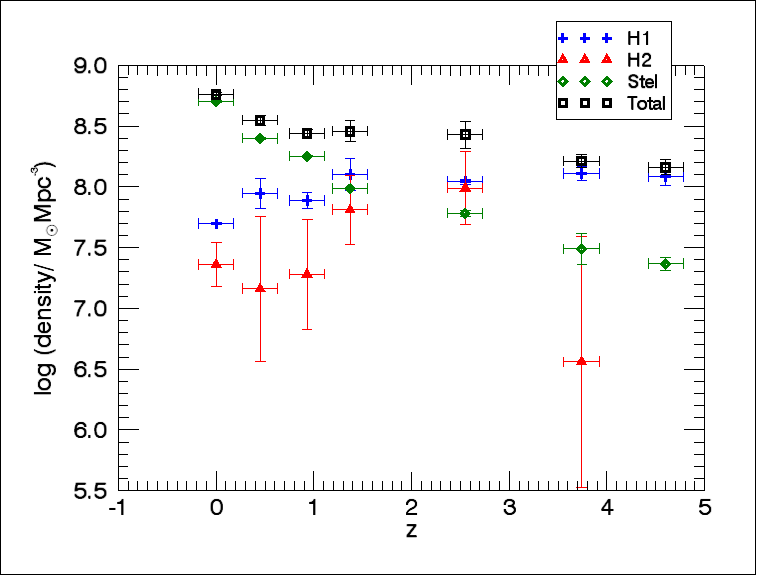}
    \caption{The gas and stellar mass densities  {\revone per unit of comoving volume} : $\rho_{HI}$ (blue crosses); $\rho_{H_{2}}$ (red triangles); $\rho_{stel}$ (green diamonds); total $\rho$, a summation of the three contributors (black squares), against redshift, z. The z `error' indicates bin width.}
    \label{fig:total_z}
\end{figure}

The binned values for all three data sets are presented in Table~\ref{tab:density} and can be seen in Figure~\ref{fig:total_z} extending between z = 0 to z = 5. The first six bins cover points where $\rho_{HI}$, $\rho_{H_{2}}$ and $\rho_{stel}$ data were all available. The seventh and final bin centred at z = 4.6 does not include $\rho_{H_{2}}$ data, as none was available at that redshift. It seemed legitimate to continue to seven bins, as the contribution to $\rho_{tot}$ from $\rho_{H_{2}}$ was already very minimal in the sixth bin, relative to those of $\rho_{HI}$ and $\rho_{stel}$. There is a clearly decreasing trend in $\rho_{tot}$. This is primarily due to the $\rho_{stel}$  data dominating at z < 2. 

Both GAD and GARD are presented in Table~\ref{tab:GAD&GARD}. The value of GAD (see Equation~\ref{eq:rho_acc}) represents the density present at a lower redshift that had not been present previously, either as a fresh gas reservoir or newly formed star. For plots of GAD against redshift, z see Appendix~\ref{sec:further_plots}. Figure~\ref{fig:GARD_SFR_z} compares the GARD with SFRD. It can be seen that the GARD values seem to by relatively constant. This does not tally with the clear drop in SFRD observed from between z = 1.5 to 0.

{\revone The errors in GAD and GARD were calculated by propagating the errors of individual densities in a standard manner,}
{\revtwo whereas the errors of these densities are the errors of the means of the densities in relevant redshift ranges reported in other studies.}

\begin{figure}
	\includegraphics[width=\columnwidth]{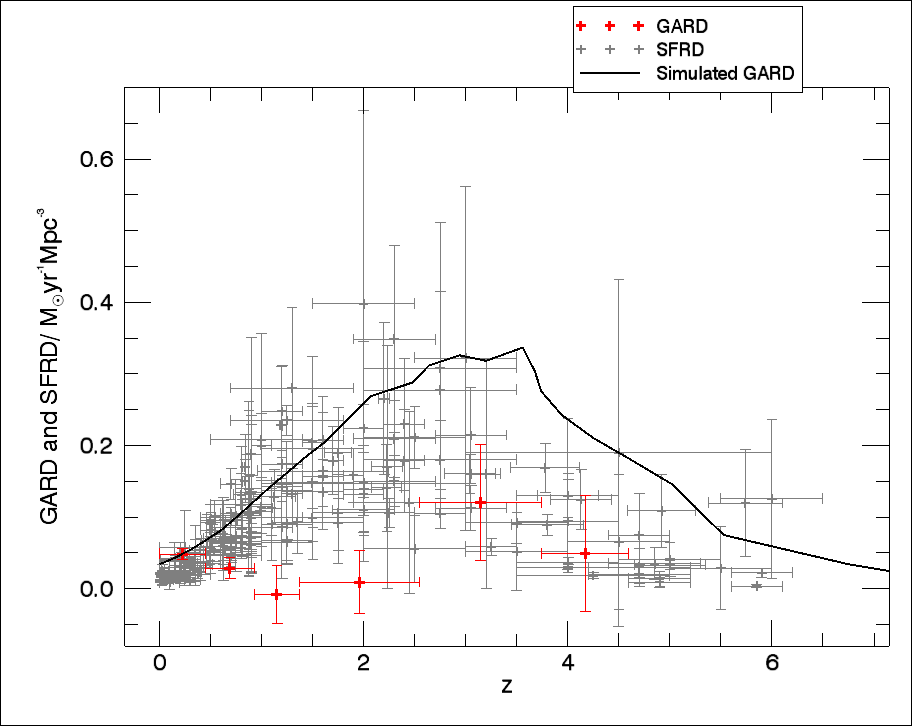}
    \caption{SFRD (grey crosses) using a compilation from~\citet{Michalowski10}, the GARD (this work, red symbols), and the {\revone simulated GARD for smooth gas accretion~\citep[fig. 12 of][]{Keres05}}. {\revtwo Due to a slight increase in $\rho_{tot}$ between the bins centred at z = 1.37 and 0.93 in an otherwise downward trend (see Table~\ref{tab:density}) the GAD value --- and therefore the GARD value --- spanning these bins is negative.} For the GAD values, see Table~\ref{tab:GAD&GARD}.}
    \label{fig:GARD_SFR_z}
\end{figure}

	\begin{table}
		\centering
		\caption{Binned comoving gas and stellar mass densities.}
		\begin{tabular}{| c | c | c | c | c |}
			\hline
			z & log($\rho_{HI}$)  & log($\rho_{H_{2}}$)  & log($\rho_{stel}$) & log($\rho_{tot}$)\\
			 &  [$M_{\odot}Mpc^{-3}$] & [$M_{\odot}Mpc^{-3}$] & [$M_{\odot}Mpc^{-3}$] & [$M_{\odot}Mpc^{-3}$]\\
			\hline
			0.0  & 7.70 $\pm$ 0.01 & 7.36 $\pm$ 0.18 & 8.70 $\pm$ 0.01 & 8.76 $\pm$ 0.01\\
			0.45 & 7.95 $\pm$ 0.12 & 7.16 $\pm$ 0.60 & 8.40 $\pm$ 0.01 & 8.55 $\pm$ 0.04\\
			0.93 & 7.89 $\pm$ 0.06 & 7.28 $\pm$ 0.45 & 8.25 $\pm$ 0.01 & 8.44 $\pm$ 0.04\\
			1.37 & 8.10 $\pm$ 0.13 & 7.81 $\pm$ 0.28 & 7.98 $\pm$ 0.01 & 8.46 $\pm$ 0.09\\
			2.55 & 8.04 $\pm$ 0.02 & 7.99 $\pm$ 0.30 & 7.78 $\pm$ 0.03 & 8.43 $\pm$ 0.12\\
			3.74 & 8.11 $\pm$ 0.06 & 6.56 $\pm$ 1.03 & 7.49 $\pm$ 0.13 & 8.21 $\pm$ 0.06\\
			4.60 & 8.09 $\pm$ 0.07 &  --- & 7.37 $\pm$ 0.05 & 8.16 $\pm$ 0.06\\
			\hline

		\end{tabular}
		\label{tab:density}
	\end{table}

	\begin{table}
		\centering
		\caption{Gas accretion density (GAD) and gas accretion rate density (GARD). {\revone The time spanned by each of the redshift bins is included under $\Delta\,time$}.}
		\begin{tabular}{| c | c | c | c |}
			\hline
			z & {$\Delta\,time$} & {\revtwo GAD} & {\revtwo GARD} \\
			&  {[Gyr]} & [$10^{7}\,M_{\odot}Mpc^{-3}$] & [$10^{-2}\,M_{\odot}yr^{-1}Mpc^{-3}$]\\
			\hline
			0.45 - 0.0  & {4.76} & {\revtwo $22.59\,\pm\,3.51$} & {\revtwo $4.83\,\pm\,0.75$}\\
			0.93 - 0.45 & {2.75} & {\revtwo $7.88\,\pm\,3.96$} & {\revtwo $2.86\,\pm\,1.44$}\\
			1.37 - 0.93 & {1.51} & {\revtwo $-1.25\,\pm\,6.07$} & {\revtwo $-0.83\,\pm\,4.03$}\\
			2.55 - 1.37 & {2.01} & {\revtwo $1.81\,\pm\,8.81$} & {\revtwo $0.90\,\pm\,4.38$}\\
			3.74 - 2.55 & {0.88} & {\revtwo $10.56\,\pm\,7.11$} & {\revtwo $12.02\,\pm\,8.09$}\\
			4.60 - 3.74 & {0.36} & {\revtwo $1.78\,\pm\,2.91$} & {\revtwo $4.92\,\pm\,8.06$}\\
			\hline

		\end{tabular}
		\label{tab:GAD&GARD}
	\end{table}
	
\section{Discussion}

{\revthree We did not find significant variations in GARD. Indeed, a straight line fit to the GARD vs.~redshift plot resulted in a slope of $-0.009\pm0.013\,M_{\odot}yr^{-1}Mpc^{-3}$, consistent with no evolution. However, argument could be made for an increasing trend in GARD from z = 2 to the present}. This is an unexpected result as it does not tally up with the SFRD which increases relatively steadily from z = 5 to z = 1.5, and then declines from around z = 1.5 to z = 0. Therefore, the decline in SFRD observed cannot be attributed to a change in gas accretion rate. It seems that galaxies in the earlier universe used up their gas supplies faster than the accretion of fresh gas could maintain (SFRD > GARD). SFRD now has dropped beneath the GARD, so the current SFRD is now sustainable. This drop in SFRD is therefore not due to decrease in gas supply. It could be because average density of gas in galaxies dropped, leaving significant amounts of gas below the star-formation threshold. 

{\revone In order to compare our unexpected results with those from a well established simulation} Figure~\ref{fig:GARD_SFR_z} compares GARD from this work with that obtained from a {\revone smoothed particle hydrodynamics simulation {\revtwo including dark matter, gas, and stars} from~\citet{Keres05}}. Simulated GARD shows a distinctive decline from z < 4 which more closely echoes SFRD than the GARD from this work, however, there is a convincing order of magnitude consistency between the values. There are a few feasible explanations for the discrepancies between the simulation and the results from this work. It is possible that the {\sc Hi} densities based on DLA data led to underestimations at higher redshifts for GARD. It is also possible that the simulation did not properly account for gas which inflows but is then expelled shortly after, so is not present inside galaxies at later epochs. 
{\revtwo Finally, the simulated GARD denote the gas accreted onto dark matter halos, whereas we measure gas accretion onto galaxies, as explained in Section~\ref{sec:method}. Assuming that some gas accretes onto halos, but does not end up inside galaxies, it is not surprising that the simulated GARD is higher than the measured one.}

This work could be considerably improved upon by a wider range of data. The lack of $\mbox{H}_2$ data at higher redshifts might have affected the gas accretion determination. This will be improved by the Atacama Large Millimeter Array (ALMA) and the Northern Extended Millimeter Array (NOEMA) which will perform CO-line scans leading to the determination of $\rho_{H_{2}}$~\citep{Walter14, Walter16}. On the other hand the Square Kilometre Array (SKA) will deliver direct measurements of $\rho_{HI}$ at least at z < 1, which will make it possible to test the result of a constant GARD in this regime.

{\revone We do not include the ionized medium in eq.~\ref{eq:rho_total} because, even though its filling factor is large \citep[fig.~11 of][]{Kalberla09}, its mass fraction is minor. In the Milky Way by mass the fraction of the ionized ISM ranges from $\sim2$\% \citep[from integrating ISM phase profiles in table 5 of][]{Wolfire03} to $\sim23$\% \citep[table~1.2 of][]{Draine11}. If this is common among galaxies, then the contribution of the ionized medium to the changes of the total densities we measure is therefore likely smaller than the uncertainties involved.}

\section{Conclusions}
\label{sec:conc}

Star formation rate has not been constant throughout the history of the Universe. This work sought to demonstrate whether the gas accretion onto galaxies over time has affected star formation rate. GARD in this work represents a simply calculated value that shows how much gas mass density has been accreted by galaxies over a certain redshift interval. If there was a tight correlation between GARD and SFRD then it might suggest that the rate of gas accretion is the limiting factor on SFR. However, a constant GARD, which we measured, might suggest that factors other than the quantity of gas being accreted affect SFRD.

\section*{Acknowledgements}

We thank our referee Jorge S\'anchez Almeida for useful comments, and Avery Meiksin, Jochan Fynbo, and Sijing Shen for advise on ISM and DLAs.
M.J.M.~acknowledges the support of the UK Science and Technology Facilities Council (STFC) and
the National Science Centre, Poland through the POLONEZ grant 2015/19/P/ST9/04010.  This project has received funding from the European Union's Horizon 2020 research and innovation programme under the Marie Sk{\l}odowska-Curie grant agreement No. 665778. 

%%%%%%%%%%%%%%%%%%%%%%%%%%%%%%%%%%%%%%%%%%%%%%%%%%


%%%%%%%%%%%%%%%%%%%%%%%%%%%%%%%%%%%%%%%%%%%%%%%%%%

%%%%%%%%%%%%%%%%% APPENDICES %%%%%%%%%%%%%%%%%%%%%

\appendix

\section{Further plots}
\label{sec:further_plots}

We plot here the complete data sets for $\rho_{HI}$ and $\rho_{stel}$, over-plotted with the binned data. The $\rho_{HI}$ data followed an increasing trend from z = 0 to 1.5, after which it seems to remain constant.

\begin{figure}
	\includegraphics[width=\columnwidth]{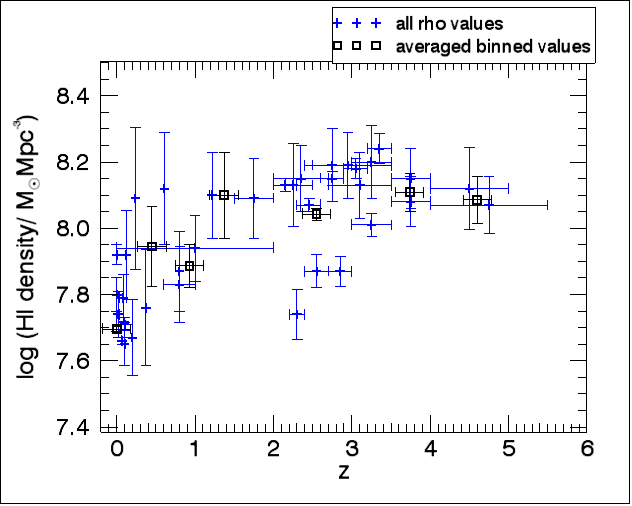}
    \caption{Atomic gas density $\rho_{HI}$ (blue crosses) vs redshift, and the average $\rho_{HI}$ for each redshift bin (black squares).}
    \label{fig:HI_z}
\end{figure}

\begin{figure}
	\includegraphics[width=\columnwidth]{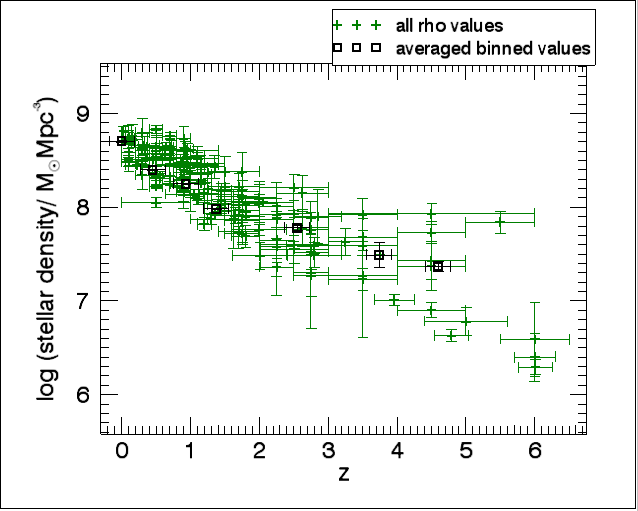}
    \caption{Stellar mass density $\rho_{stel}$ (green crosses) vs redshift, and the average $\rho_{stel}$ for each redshift bin (black squares).}
    \label{fig:stel_z}
\end{figure}

\begin{figure}
	\includegraphics[width=\columnwidth]{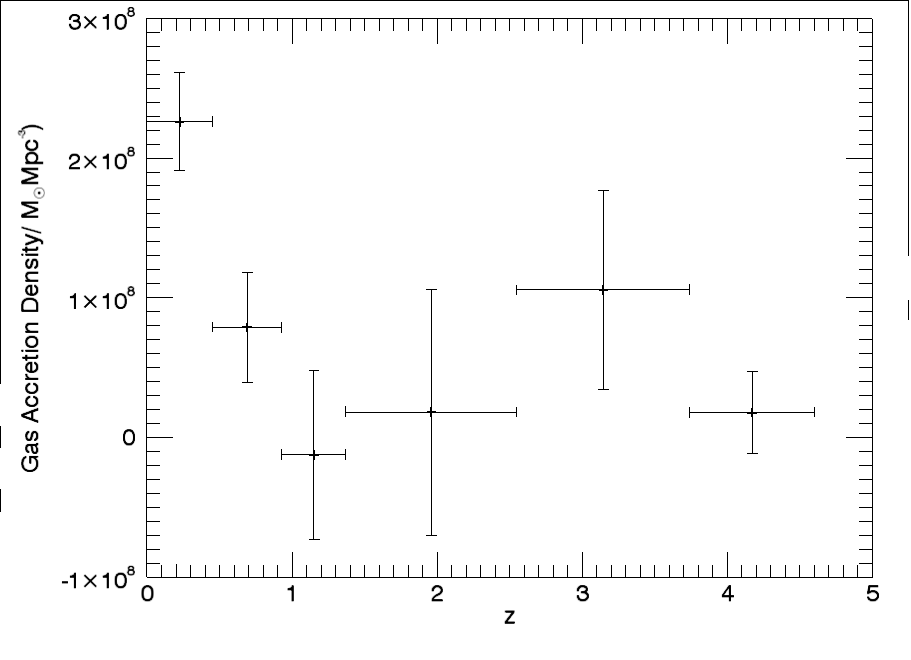}
    \caption{Gas Accretion Density (GAD) against redshift, z. {\revtwo Due to a slight increase in $\rho_{tot}$ between the bins centred at z = 1.37 and 0.93 in an otherwise downward trend (see Table~\ref{tab:density}) the GAD value spanning these bins is negative.}}
    \label{fig:gad_z}
\end{figure}

\section{$\rho_{HI}$ Data Compilation}

 The $\rho_{H_{2}}$ data came from~\citet{Keres03} and ~\citet{Decarli16}. The full $\rho_{stel}$ compilation was sourced from~\citet{Michalowski10}. Our complete $\rho_{HI}$ compilation is presented in full in table~\ref{tab:HIdata}.

	\begin{table*}
		\centering
		\caption{Compilation of atomic gas mass densities, $\rho_{HI}$. {DLA indicates damped Lyman-$\alpha$.} }
		\begin{tabular}{| c | c | c | c | c | c | c |}
			\hline
			z & z width & log $\rho_{HI}$ & lower error log $\rho_{HI}$ & upper error log $\rho_{HI}$ & {Type of $\rho_{HI}$ }& Reference \\
			 &  & [$M_{\odot}Mpc^{-3}$] & [$M_{\odot}Mpc^{-3}$] & [$M_{\odot}Mpc^{-3}$] & {measurement} & \\
			\hline
			0.0 & 0.0 & 7.80 & 0.05 & 0.05 & {{\sc Hi}} & ~\citet{Zwaan05} \\
			0.0 & 0.0 & 7.92 & 0.03 & 0.03 & {{\sc Hi}} & ~\citet{Martin10} \\
			0.125 & 0.0 & 7.92 & 0.15 & 0.12 & {{\sc Hi}} & ~\citet{Freudling11} \\
			0.065 & 0.0 & 7.66 & 0.01 & 0.01 & {{\sc Hi}} & ~\citet{Hoppmann15} \\
			0.1 & 0.1 & 7.71 & 0.02 & 0.02 & {{\sc Hi}} & ~\citet{Hoppmann15} \\
			0.24 & 0.0 & 8.09 & 0.27 & 0.16 & {{\sc Hi} stacking} & ~\citet{Lah07} \\
			0.02 & 0.02 & 7.74 & 0.10 & 0.04 & {{\sc Hi} stacking} & ~\citet{Delhaize13} \\
			0.085 & 0.045 & 7.79 & 0.09 & 0.05 & {{\sc Hi} stacking} & ~\citet{Delhaize13} \\
			0.1 & 0.0 & 7.65 & 0.07 & 0.06 & {{\sc Hi} stacking} & ~\citet{Rhee13} \\
			0.2 & 0.0 & 7.67 & 0.13 & 0.10 & {{\sc Hi} stacking} & ~\citet{Rhee13} \\
			0.37 & 0.0 & 7.76 & 0.21 & 0.14 & {{\sc Hi} stacking} & ~\citet{Rhee16} \\
			0.8 & 0.0 & 7.87 & 0.14 & 0.10 & {{\sc Hi} intensity} & ~\citet{Chang10} \\
			&&&&&{mapping} & \\
			0.8 & 0.2 & 7.83 & 0.13 & 0.10 & {{\sc Hi} intensity} & ~\citet{Masui13} \\
			&&&&&{mapping} & \\
			1.0 & 1.0 & 7.94 & 0.10 & 0.10 & {DLA} & ~\citet{Peroux03} \\
			2.35 & 0.35 & 8.15 & 0.10 & 0.10 & {DLA} & ~\citet{Peroux03} \\
			3.1 & 0.4 & 8.13 & 0.10 & 0.10 & {DLA} & ~\citet{Peroux03} \\
			2.95 & 0.55 & 8.19 & 0.10 & 0.10 & {DLA} & ~\citet{Peroux03} \\
			0.609 & 0.0 & 8.12 & 0.20 & 0.14 & {DLA} & ~\citet{Rao06} \\
			1.219 & 0.0 & 8.10 & 0.15 & 0.11 & {DLA} & ~\citet{Rao06} \\
			2.15 & 0.15 & 8.13 & 0.02 & 0.02 & {DLA} & ~\citet{Noterdaeme12} \\
			2.45 & 0.15 & 8.07 & 0.02 & 0.02 & {DLA} & ~\citet{Noterdaeme12} \\
			2.75 & 0.15 & 8.15 & 0.02 & 0.02 & {DLA}  & ~\citet{Noterdaeme12} \\
			3.05 & 0.15 & 8.18 & 0.03 & 0.03 & {DLA} & ~\citet{Noterdaeme12} \\
			3.35 & 0.15 & 8.24 & 0.05 & 0.04 & {DLA} & ~\citet{Noterdaeme12} \\
			1.75 & 0.25 & 8.09 & 0.14 & 0.10 & {DLA} & ~\citet{Zafar13} \\
			2.25 & 0.25 & 8.13 & 0.14 & 0.11 & {DLA} & ~\citet{Zafar13} \\
			2.75 & 0.25 & 8.19 & 0.12 & 0.10 & {DLA} & ~\citet{Zafar13} \\
			3.25 & 0.25 & 8.20 & 0.12 & 0.10 & {DLA} & ~\citet{Zafar13} \\
			3.75 & 0.25 & 8.15 & 0.10 & 0.08 & {DLA} & ~\citet{Zafar13} \\
			4.5 & 0.5 & 8.12 & 0.14 & 0.11 & {DLA} & ~\citet{Zafar13} \\
			2.3 & 0.1 & 7.74 & 0.08 & 0.07 & {DLA} & ~\citet{Prochaska09} \\
			2.55 & 0.15 & 7.87 & 0.05 & 0.05 & {DLA} & ~\citet{Prochaska09} \\
			2.85 & 0.15 & 7.87 & 0.05 & 0.04 & {DLA} & ~\citet{Prochaska09} \\
			3.25 & 0.25 & 8.01 & 0.04 & 0.03 & {DLA} & ~\citet{Prochaska09} \\
			3.75 & 0.25 & 8.08 & 0.08 & 0.07 & {DLA} & ~\citet{Prochaska09} \\
			4.75 & 0.75 & 8.07 & 0.09 & 0.08 & {DLA} & ~\citet{Prochaska09} \\	
			\hline
		\end{tabular}
		\label{tab:HIdata}
	\end{table*}

%%%%%%%%%%%%%%%%%%%%%%%%%%%%%%%%%%%%%%%%%%%%%%%%%%

% Don't change these lines
\bsp	% typesetting comment
\label{lastpage}
\end{document}